\documentstyle[aps,prb,preprint,floats]{revtex}
\input epsf
\begin{document}
\draft
\preprint{UCSBTH--94--18}
\title{{\bf
Neutron scattering in a \bbox{d_{\bf x^2-y^2}}--wave superconductor
with strong impurity scattering and Coulomb correlations}}
\author{\large S.~M.~Quinlan and D.~J.~Scalapino}
\address{Department of Physics,
University of California,
Santa Barbara, CA 93106--9530}
\date{June 16, 1994}
\maketitle
\begin{abstract}
We calculate the spin susceptibility at and below $T_c$ for a
$d_{x^2-y^2}$--wave superconductor with resonant impurity scattering
and Coulomb correlations. Both the impurity scattering and the Coulomb
correlations act to maintain peaks in the spin susceptibility, as a
function of momentum, at the Brillouin zone edge.  These peaks would
otherwise be suppressed by the superconducting gap.  The predicted
amount of suppression of the spin susceptibility in the
superconducting state compared to the normal state is in qualitative
agreement with results from recent magnetic neutron scattering
experiments on La$_{1.86}$Sr$_{0.14}$CuO$_4$ for momentum values at
the zone edge and along the zone diagonal.  The predicted peak widths
in the superconducting state, however, are narrower than those in the
normal state, a narrowing which has not been observed experimentally.
\end{abstract}

\narrowtext

\section{Introduction}

\begin{figure}
\epsfxsize=468bp
\epsfbox{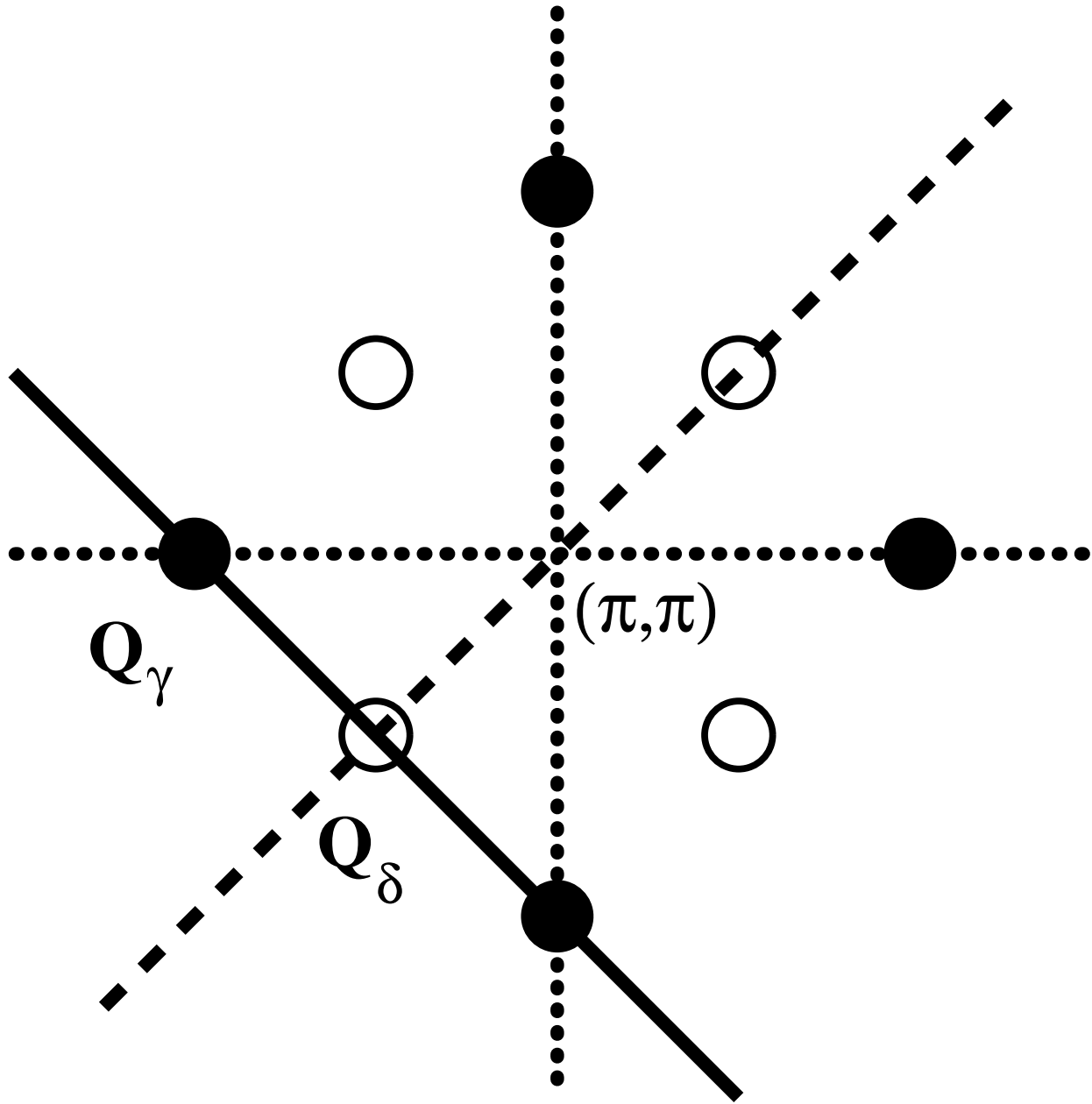}
\vspace{36bp}
\caption{Map of reciprocal space probed by
the neutron scattering experiments. Dotted lines represent the
Brillouin zone edges near the $(\pi,\pi)$ point. The dashed line
indicates the zone diagonal. Filled circles show the locations of
neutron scattering intensity peaks in the normal state. Open circles
represent the expected positions for scattering intensity peaks for a
clean $d$--wave superconductor well below $T_c$. The solid line shows
the direction of the momentum scans shown in
Fig.~\protect\ref{pureVSimp} and
Figs.~\protect\ref{vertVSnovert}-\protect\ref{momentavg}.}
\label{peaklayout}
\end{figure}

Recent magnetic neutron scattering experiments on
La$_{1.86}$Sr$_{0.14}$CuO$_4$ by Mason and co-workers\cite{Mason} show
an isotropic but incomplete suppression of the scattering intensity
below $T_c$ for small energy transfers and momentum transfers near the
$(\pi,\pi)$ point of the Brillouin zone.  In the normal state, the
magnetic neutron scattering intensity, as a function of momentum,
exhibits four peaks displaced slightly along the edges of the
Brillouin zone away from the $(\pi,\pi)$ point at the corner of the
zone, as shown in Fig.~\ref{peaklayout}.
The experimental results
show a suppression of these peaks, compared to their magnitude at
$T_c$, of about 60\% at a temperature of $0.13T_c$.

For an isotropic $s$--wave superconductor we would expect an isotropic
suppression of this scattering intensity as the temperature drops
below $T_c$. But for a clean $s$--wave state at low temperatures the
intensity should die away exponentially with temperature. The moderate
suppression shown by experiment suggests that if $s$--wave
superconductivity is present, it is gapless and isotropic.

Alternatively, a superconductor with $d_{x^2-y^2}$ symmetry is always
gapless, so an incomplete suppression of the scattering intensity is
to be expected.  However, for a clean $d$--wave superconductor it is
expected that this suppression will not be isotropic. Lu\cite{Lu} and
Zha and co-workers\cite{ZhaLevinSi} have pointed out that in the
superconducting state the four normal state peaks should be
suppressed, leaving only the response due to particle-hole excitations
associated with the nodes. This response is displaced slightly away
from $(\pi,\pi)$ along the zone diagonals, giving a low temperature
structure which appears to have undergone a 45 degree rotation about
$(\pi,\pi)$.  The failure to observe this rotation has raised
questions regarding whether a $d_{x^2-y^2}$ description of the
superconducting state of La$_{2-x}$Sr$_x$CuO$_4$ is viable.

Now, for $d_{x^2-y^2}$ pairing it is known that resonant impurity
scattering leads to important effects for both the penetration
depth\cite{HirschfeldGoldenfeld} and the NMR response.\cite{kitaoka}
In addition, the effects of the Coulomb enhanced spin fluctuations are
believed to play an essential role in the cuprates. As previously
discussed,\cite{NejatXrpa,Maki} when the effects of Coulomb
correlations on the magnetic neutron scattering are included within
the RPA approximation, the $d_{x^2-y^2}$ results can be brought into
closer agreement with the experimental observations.

Here we investigate both of these effects by calculating the neutron
scattering for a model of a dirty $d_{x^2-y^2}$--wave superconductor
with strong Coulomb correlations.  Specifically, we consider a Hubbard
model on a two-dimensional square lattice with a near-neighbor hopping
$t$ and an on-site Coulomb interaction $U$.  To this we add impurity
scattering in the form of a dilute random array of zero-range
scattering potentials $V$.

The magnetic neutron scattering intensity is given by a structure
factor $S({\bf q},\omega) = [n(\omega)+1] \chi''({\bf q},\omega)$,
which is proportional to the imaginary part of the spin susceptibility
$\chi({\bf q},\omega)$. We calculate $\chi({\bf q},\omega)$ in two
stages.  First, in Sec.~\ref{impuritysection}, we examine the effects
of the impurities on $\chi({\bf q},\omega)$, including self-energy as
well as vertex corrections. Then, in Sec.~\ref{coulombsection}, the
strong spin susceptibility enhancement effects of the Coulomb
interaction are included within an RPA-type approximation.  Section
\ref{conclusions} contains our conclusions.

Before beginning, it is useful to review the behavior expected for a
system described by a two-dimensional tight-binding band which
undergoes a transition from a normal state to a $d_{x^2-y^2}$ BCS
superconducting state in the absence of impurity scattering and strong
Coulomb correlations.  As discussed in Littlewood {\it et
al.},\cite{Littlewood} the four normal state peaks shown in
Fig.~\ref{peaklayout} arise due to a combination of the weak
requirements for favorable nesting in two dimensions combined with
umklapp scattering processes. As a $d$--wave gaps opens up below $T_c$
the thermally available scattering states are restricted to those near
the nodes of the $d$--wave gap function and these peaks are
suppressed. The favorable low-energy momentum transfers are then
restricted to those which represent node-to-node scattering, and the
corresponding wavevectors lie along the zone diagonal directions. Thus
at low temperatures and energies well below $2\Delta(0)$, peaks in the
neutron scattering should be observed to lie along the zone
diagonals.\cite{Lu,ZhaLevinSi}

Here we find that impurity scattering and Coulomb correlations both
act to keep the scattering intensity peaks at their normal state
positions along the Brillouin zone edge. For reasonable parameters,
the amount of suppression of the scattering intensity in the
dirty--$d_{x^2-y^2}$ superconducting state compared to the normal
state is consistent with the experimental observations both at the
zone edge and along the diagonal. However, the predicted peak widths
in the superconducting state are narrower than those in the normal
state, a narrowing which has not been observed experimentally.

\section{Impurity scattering effects}
\label{impuritysection}

In this section we examine the effect of impurities on the spin
susceptibility of a BCS $d_{x^2-y^2}$ state. We begin by including the
self-energy effects and then consider the impurity vertex corrections.

\subsection{Self-energy corrections}
\label{selfenergysection}

\begin{figure}
\epsfxsize=468bp
\epsfbox{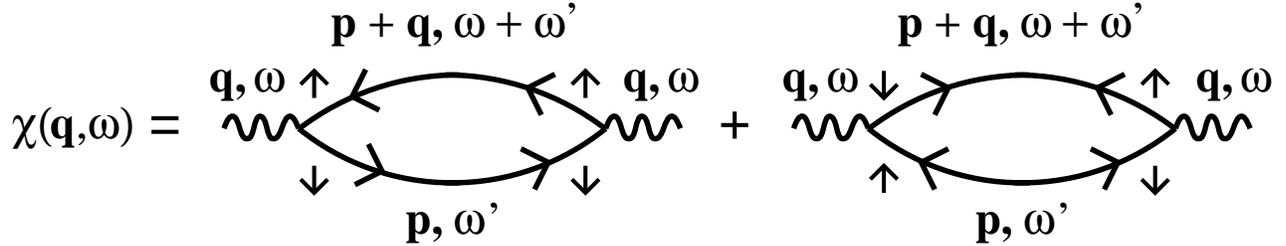}
\vspace{36bp}
\caption{Lowest order bubble diagram representing the spin
susceptibility probed in magnetic neutron scattering experiments.}
\label{bubble}
\end{figure}

Neglecting the vertex corrections, the spin susceptibility in the
superconducting state is given by the particle-hole bubble diagrams
shown in Fig.~\ref{bubble}.
The bare vertex representing magnetic
neutron scattering is understood to connect only electrons of opposite
spin. Evaluating this diagram, continuing to real frequencies, and
taking the imaginary part, we obtain the following formula for
$\chi''({\bf q},\omega)$:

\begin{eqnarray}
\chi''({\bf q},\omega) = -\pi \int {d^2p \over (2 \pi)^2} ~ d\omega' ~
  [f(\omega+\omega') - f(\omega')] ~
  [&&A_N({\bf p}+{\bf q},\omega+\omega') A_N({\bf p},\omega') \nonumber\\
  &&+ A_A({\bf p}+{\bf q},\omega+\omega') A_A({\bf p},\omega')].
\end{eqnarray}
$A_N({\bf p},\omega)$ and $A_A({\bf p},\omega)$ are the spectral
weights of the normal and anomalous electron propagators,
respectively:

\begin{mathletters}
\begin{eqnarray}
A_N({\bf p},\omega) = - {1 \over \pi} ~ Im ~ G({\bf p},\omega), \\
\nonumber\\
A_A({\bf p},\omega) = - {1 \over \pi} ~ Im ~ F({\bf p},\omega),
\end{eqnarray}
\end{mathletters}
where
\begin{mathletters}
\begin{eqnarray}
G({\bf p},\omega) = {\tilde \omega + \tilde \varepsilon_{\bf p}
\over \tilde \omega^2 -
\tilde \varepsilon_{\bf p}^2 - \tilde \Delta_{\bf p}^2}
\end{eqnarray}
and
\begin{eqnarray}
F({\bf p},\omega) = {-\tilde \Delta_{\bf p}
\over \tilde \omega^2 -
\tilde \varepsilon_{\bf p}^2 - \tilde \Delta_{\bf p}^2}.
\end{eqnarray}
\end{mathletters}
The tilde symbol indicates inclusion of the impurity scattering
self-energy corrections:

\begin{mathletters}
\begin{eqnarray}
\tilde \omega &&= \omega-\Sigma_0, \\
\tilde \varepsilon_{\bf p} &&= \varepsilon_{\bf p} + \Sigma_3, \\
\tilde \Delta_{\bf p} &&= \Delta_{\bf p} + \Sigma_1.
\end{eqnarray}
\end{mathletters}
Here we are using $\varepsilon_{\bf p} = -2t(\cos p_x +\cos p_y)-\mu$ and
$\Delta_{\bf p} = a\Delta(T)\,(\cos p_x - \cos p_y)$. The parameter
$a$ is chosen such that the maximum value of the gap on the Fermi
surface is $\Delta(T)$.

The effect of impurity scattering may now be included by allowing the
electron self energy to include multiple scattering from the
potentials $V$.
\begin{figure}
\epsfxsize=468bp
\epsfbox{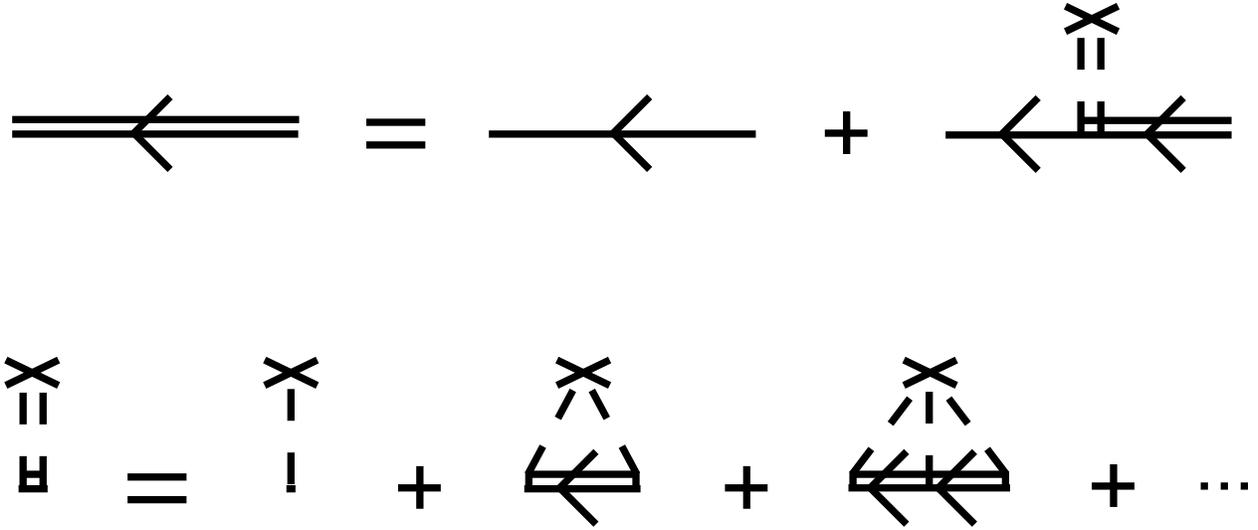}
\vspace{36bp}
\caption{Diagrammatic representation of the calculation of the
electron self energy due to impurity scattering in the non-crossing or
dilute limit.}
\label{tmatrix}
\end{figure}
Figure \ref{tmatrix} shows these scattering processes
for the dilute or non-crossing limit. Electrons are allowed to scatter
multiply from an impurity to allow for arbitrarily strong scattering
strengths. In the dilute limit, however, the electrons are assumed to
interact with only one impurity at a time so the impurity interaction
lines may not cross. The self energy within this approximation is
given by\cite{hirschfeld}

\begin{mathletters}
\label{selfenergy}
\begin{eqnarray}
\Sigma_0(\omega) = \Gamma_N T_0(\omega), \\
\nonumber\\
\Sigma_3(\omega) = \Gamma_N T_3(\omega),
\end{eqnarray}
\end{mathletters}
where
\begin{mathletters}
\begin{eqnarray}
T_0(\omega) = {G_0(\omega) \over {c^2 - G_0^2(\omega)}},
\end{eqnarray}
and
\begin{eqnarray}
T_3(\omega) = {-c \over {c^2 - G_0^2(\omega)}}.
\end{eqnarray}
\end{mathletters}
Here $\Gamma_N = n_i/[\pi N(0)]$, $c~=~cot~\delta_0$, and
\begin{eqnarray}
G_0(\omega) = {1 \over \pi N(0)} \int {d^2p \over (2 \pi)^2} {\tilde
\omega \over \tilde \omega^2 - \tilde \varepsilon_{\bf p}^2 - \tilde
\Delta_{\bf p}^2},
\end{eqnarray}
where $n_i$ is the impurity concentration, $N(0)$ is the normal phase
density of states, and $\delta_0$ is the scattering phase
shift.\cite{impuritynote} The self-energy correction to the gap
function $\Sigma_1$ vanishes for a $d$--wave gap, and in the unitary
limit, $c=0$, only the $\Sigma_0$ contribution remains.

\begin{figure}
\epsfxsize=396bp
\epsfbox{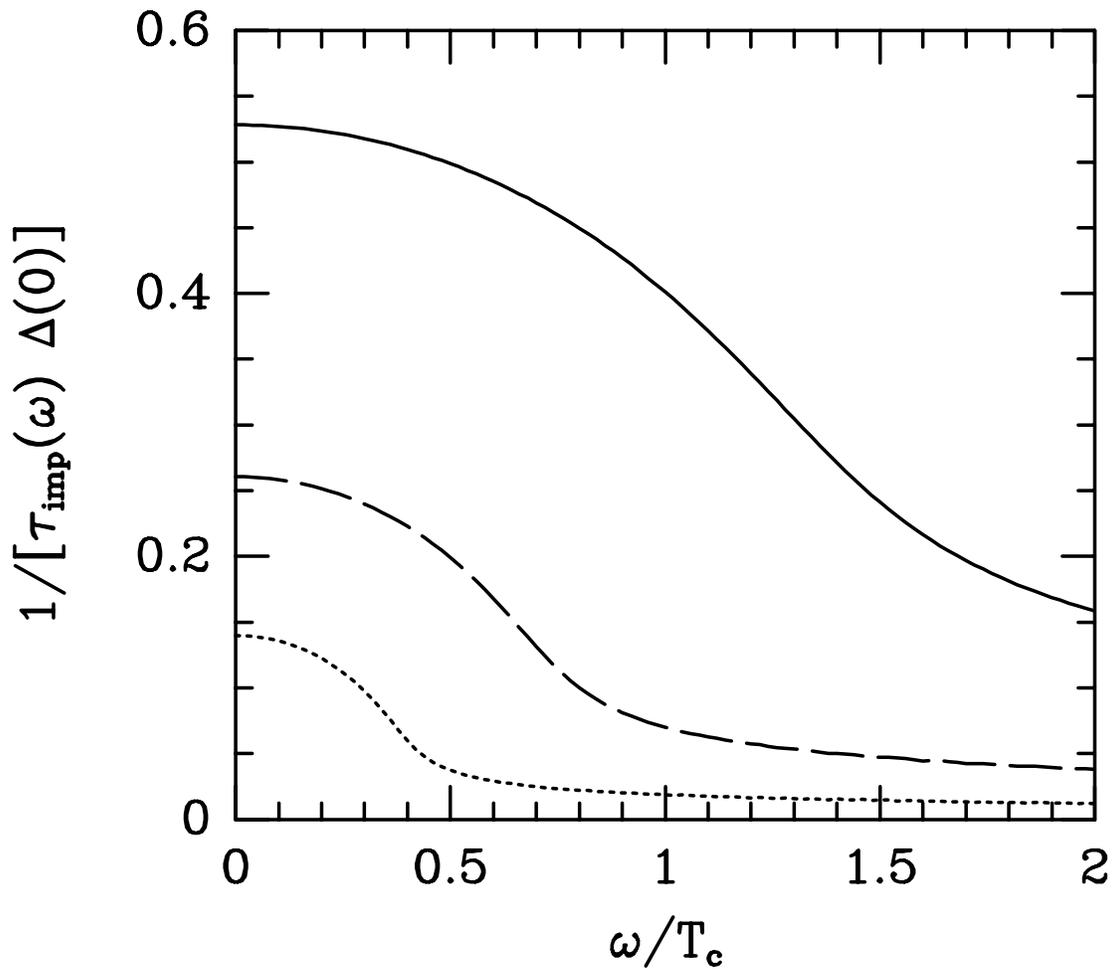}
\caption{Quasiparticle relaxation rate $1/\tau_{imp}$ due to impurity
scattering in the unitary limit, $c=0$. Results are shown for
$n_i=1\%$ (solid line), $n_i=0.3\%$ (dashed line) and $n_i=0.1\%$
(dotted line).}
\label{tauimp}
\end{figure}

We can get an idea of the behavior of the electron self energy due to
impurity scattering by looking at the quasiparticle decay rate
$1/\tau_{imp}(\omega)$ implied by the self energy:
\begin{eqnarray}
1/\tau_{imp}(\omega) = -2 Im \Sigma_0(\omega).
\end{eqnarray}
Figure \ref{tauimp} shows $1/\tau_{imp}(\omega)$ at $T=0.1T_c$ for
$n_i=1\%$, $n_i=0.3\%$, and $n_i=0.1\%$ in the resonant or unitary
limit, $c=0$. Here $\Sigma_0(\omega)$ has been evaluated for a
filling slightly below half filling $<n>=0.85$ and a low temperature
gap magnitude given by $\Delta(0)=3T_c$, where $T_c=0.05t$, parameters
which are also used in the calculations that follow.

The unitary limit is of some interest in these systems as it has been
invoked to explain the $T^2$ temperature dependence of the magnetic
penetration depth seen in the thin film cuprate
superconductors.\cite{HirschfeldGoldenfeld} In general, unitary limit
scattering leads to the largest low frequency quasiparticle decay rate
for a given impurity concentration. For very large quasiparticle decay
rates, $1/\tau_{imp}(\omega) \sim \Delta(T)$, we might expect the
impurity scattering will lead to a suppression of the superconducting
gap itself, possibly leading to a reentrant behavior of the
superconducting state. In the calculations which follow we use
$n_i=1\%$ with $c=0$, a choice of parameters such that
$1/\tau_{imp}(\omega)$ is always smaller than $\Delta(0)$.

\begin{figure}
\epsfxsize=396bp
\epsfbox{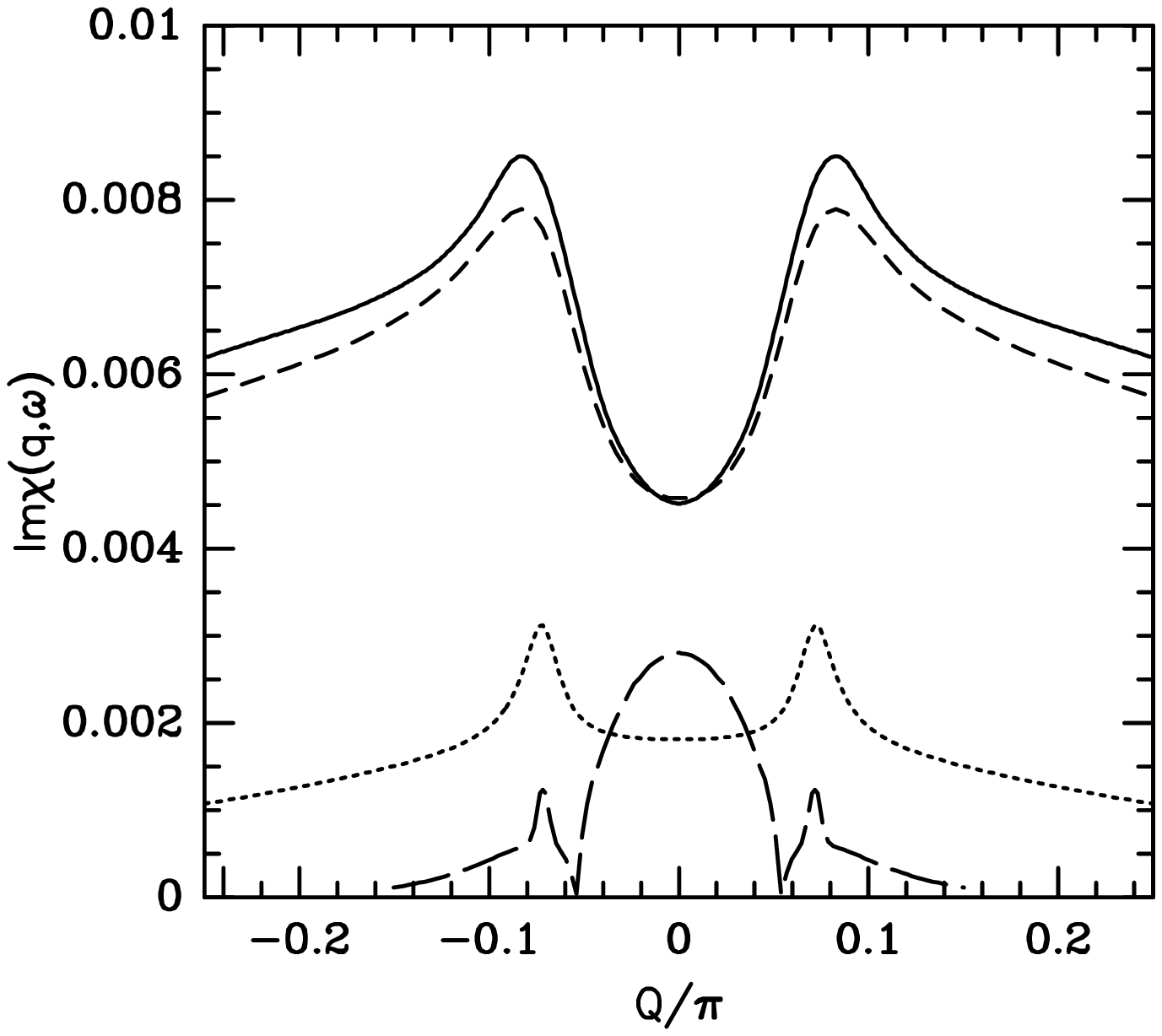}
\caption{$\chi''({\bf q},\omega)$ with and without the inclusion of
impurity self-energy effects as a function of momentum. The momentum
scan extends along the solid line in Fig.~\protect\ref{peaklayout},
perpendicular to the zone diagonal. $Q=0$ indicates the intersection
with the zone diagonal.  The vertical axis is in units of $t^{-1}$.
The curves represent the following: (solid) normal state, pure; (short
dashes) normal state, dirty; (long dashes) superconducting state,
pure; (dotted) superconducting state, dirty.}
\label{pureVSimp}
\end{figure}

Figure \ref{pureVSimp} shows the results of a calculation of
$\chi''({\bf q},\omega)$ with and without the inclusion of impurity
self-energy effects.
Each of the following figures shows $\chi''({\bf
q},\omega)$ evaluated at low frequency, $\omega=0.4T_c$, for ${\bf q}$
values which run along the solid line in Fig.~\ref{peaklayout}, a line
which passes through the normal state peaks in $\chi''({\bf
q},\omega)$. The x-axis shows the distance $Q$ in momentum space
measured from the point ${\bf Q}_\gamma=(\pi-\delta/2,\pi-\delta/2)$
where the zone diagonal (dashed line of Fig.~\ref{peaklayout})
intersects the solid line in Fig.~\ref{peaklayout}, that is, ${\bf q}
= (\pi-\delta/2+Q,\pi-\delta/2-Q)$.  For the above filling, the normal
state peaks are displaced from $(\pi,\pi)$ by an amount
$\delta=0.9\pi$. This value for $\delta$ depends upon the details of
the band structure. For a three-band Hubbard model with parameters
adjusted to fit the La$_2$CuO$_4$ band structure it can be about twice
as large\cite{ZhaLevinSi,Littlewood} in agreement with experiments on
La$_{2-x}$Sr$_x$CuO$_4$. Here for simplicity, as in some previous
studies,\cite{NejatXrpa} we use a one-band near-neighbor form for
$\varepsilon_{\bf p}$.

In Fig.~\ref{pureVSimp} curves are shown comparing the normal,
$T=T_c$, and superconducting, $T=0.1T_c$, states as well as the pure
and dirty cases. For the normal state calculation, we would expect
that the electron self energy should include contributions due to
inelastic scattering by spin fluctuations (this is discussed further
in Sec.~\ref{coulombsection}). Here instead, the self energy in the
normal state has been modeled as a frequency and momentum independent
constant, $\Sigma_0=-i\gamma$, where $\gamma=T_c/2$. It is apparent
from the figure that this self energy has only a modest effect on the
shape of the $\chi''({\bf q},\omega)$ peaks. Thus the use of such a
simple model for the self energy should not interfere with
interpretation of the experimental results.

In the superconducting state the inelastic scattering due to spin
fluctuations is suppressed by the opening of a superconducting gap.
Thus impurity scattering effects will dominate at low temperatures and
frequencies. Accordingly, the dirty superconducting state calculation
incorporates an impurity scattering self energy calculated as shown in
Eq.~(\ref{selfenergy}). Figure \ref{pureVSimp} shows the results for
an impurity concentration $n_{imp}=1\%$ with a scattering strength in
the unitary limit, $c=0$.

As may be seen in Fig.~\ref{pureVSimp}, for the pure case calculation
the zone edge peaks which occur in the normal state are greatly
suppressed in the superconducting state and a peak in $\chi''({\bf
q},\omega)$ occurs instead along the zone diagonal. Inclusion of
impurity scattering self-energy effects, however, works to restore the
zone edge peaks while the zone diagonal peak is washed out. A lack of
thermally available scattering states leads to the suppression of the
zone edge peaks for the clean superconductor. In the dirty system,
impurity scattering broadens the quasiparticle resonances in the
electron propagators allowing access to otherwise thermally restricted
scattering states and acting to restore the zone edge peaks which
arise due to scattering between such states.

\subsection{Vertex corrections}
\label{vertexsection}

\begin{figure}
\epsfxsize=468bp
\epsfbox{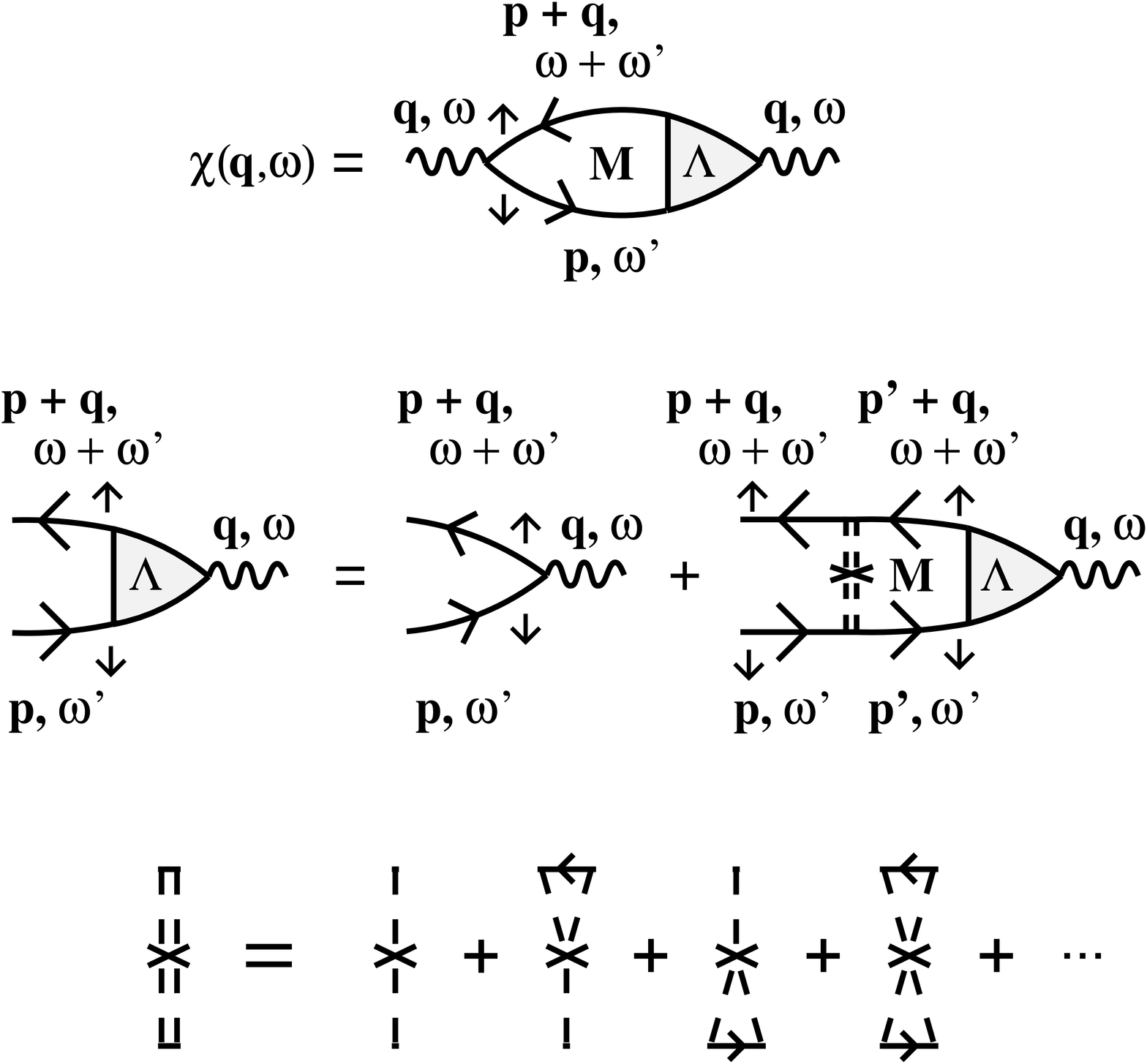}
\vspace{36bp}
\caption{Diagrammatic representation of the vertex corrections to the
spin susceptibility in the normal state due to impurity scattering in
the non-crossing or dilute limit.}
\label{vertex}
\end{figure}

The above electron self energy includes all possible non-crossing
impurity scattering contributions to the electron self energy. In
order to include all possible non-crossing impurity scattering
contributions to the spin susceptibility, we must also include vertex
corrections due to impurity scattering. That is, we must include all
processes where the electron and hole are allowed to scatter multiply
from the same impurity. The form of these vertex corrections is shown
in Fig.~\ref{vertex}.
The vertex corrected susceptibility bubble
$\chi({\bf q},i\omega_m)$ is given by a particle-hole bubble $M({\bf
q},i\omega_m,{i\omega_n})$ with a simple vertex at one end and a
dressed vertex $\Lambda({\bf q},i\omega_m,{i\omega_n})$ at the other
end. In the normal state this may be written as
\begin{eqnarray}
\chi({\bf q},i\omega_m) = T \sum_n ~ M({\bf q},i\omega_m,{i\omega_n}) ~
\Lambda({\bf q},i\omega_m,{i\omega_n}),
\label{Xvertnorm}
\end{eqnarray}
where
\begin{eqnarray}
M({\bf q},i\omega_m,{i\omega_n}) = - \int {d^2p \over (2 \pi)^2} ~
G({\bf p}+{\bf q},{i\omega_m}+{i\omega_n}) ~ G({\bf p},{i\omega_n}).
\end{eqnarray}
The dressed vertex is given by the sum of a series of ladder diagrams
where impurity scattering lines [represented by
$\Gamma({i\omega_m},{i\omega_n})$] are allowed to connect the particle
and hole lines:
\begin{eqnarray}
\Lambda({\bf q},i\omega_m,{i\omega_n}) = 1 +
\Gamma({i\omega_m},{i\omega_n}) ~ M({\bf q},i\omega_m,{i\omega_n}) ~
\Lambda({\bf q},i\omega_m,{i\omega_n}).
\label{lambdanorm}
\end{eqnarray}
Equations (\ref{Xvertnorm}) and (\ref{lambdanorm}) are easily rearranged
to yield
\begin{eqnarray}
\chi({\bf q},i\omega_m) = T \sum_n ~ {M({\bf q},i\omega_m,{i\omega_n})
\over 1 - \Gamma({i\omega_m},{i\omega_n}) ~ M({\bf q},i\omega_m,{i\omega_n})}.
\end{eqnarray}

Since the impurity scattering lines include any number of multiple
scattering events, they are given simply by a product of two of the
t-matrices used in the self energy calculation:
\begin{eqnarray}
\Gamma({i\omega_m},{i\omega_n}) = - {n_i \over [\pi N(0)]^2} ~
T_0({i\omega_m}+{i\omega_n}) ~ T_0({i\omega_n}).
\end{eqnarray}
Since we are considering only the unitary scattering limit, i.e.
$c=0$, the $T_3$ component of the scattering t-matrix vanishes. Thus
$\Gamma({i\omega_m},{i\omega_n})$ contains contributions from $T_0$
only.

\begin{figure}
\epsfxsize=468bp
\epsfbox{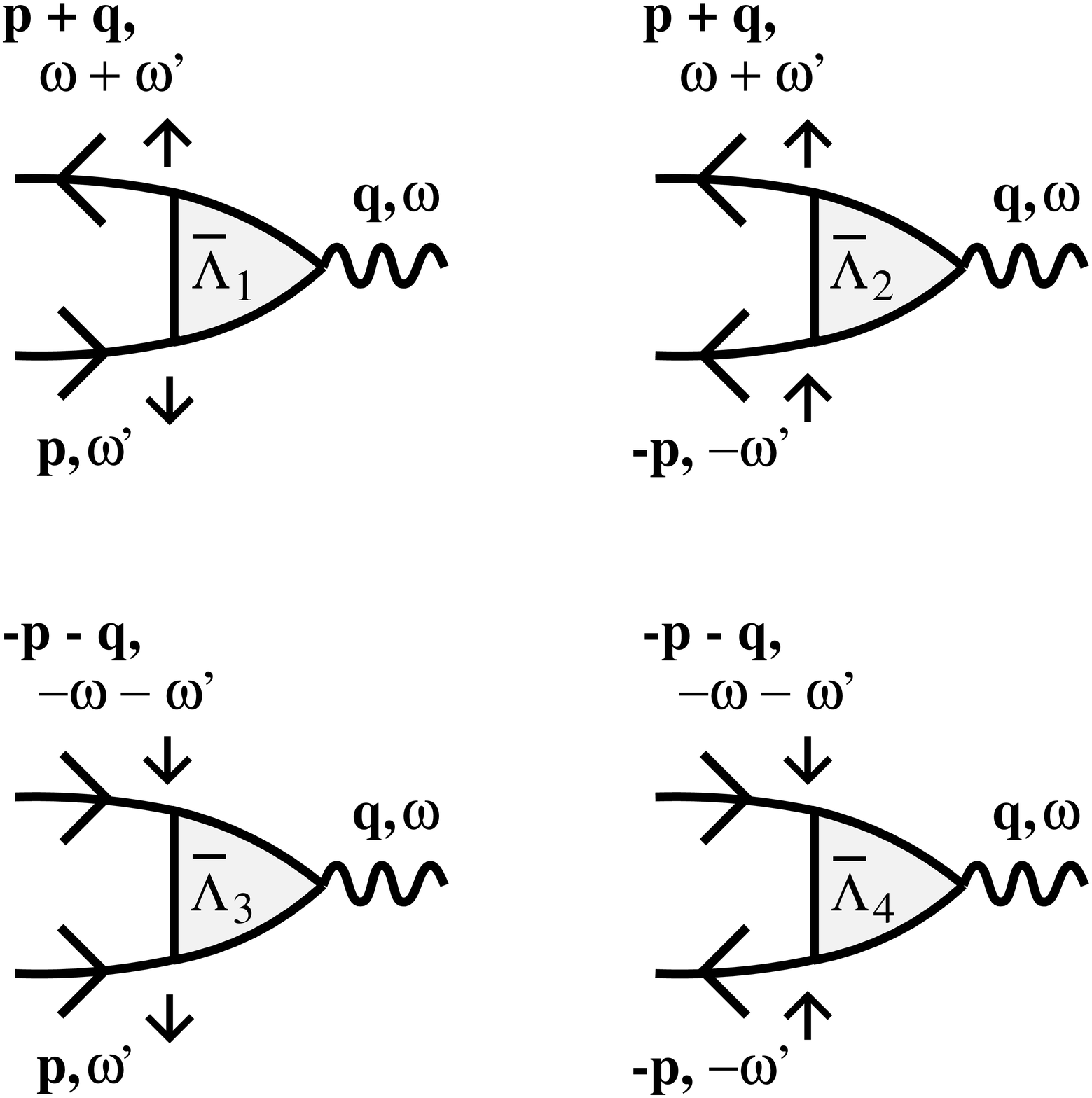}
\vspace{36bp}
\caption{Components of the dressed vertex in the superconducting state.}
\label{vertcomponents}
\end{figure}

\begin{figure}
\epsfxsize=468bp
\epsfbox{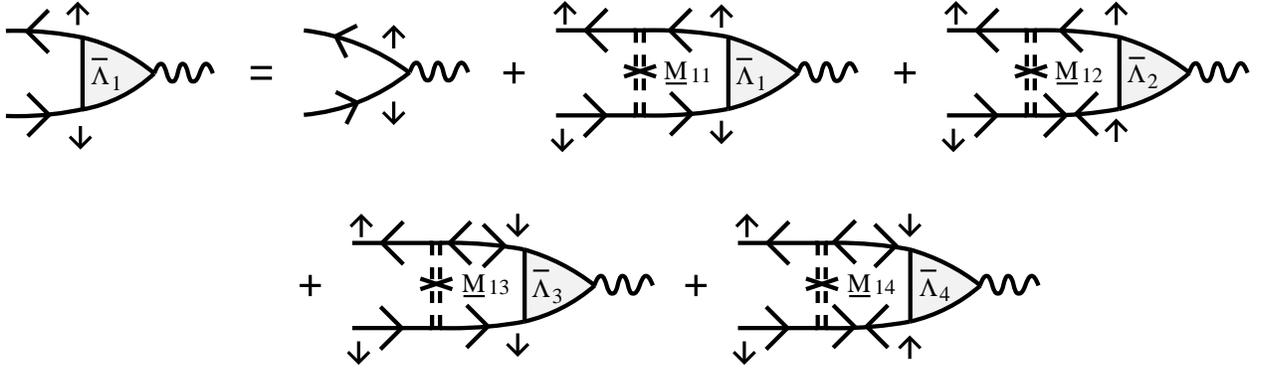}
\vspace{36bp}
\caption{Diagrammatic representation of the first component of the
matrix equation, Eq.~(\protect\ref{lambdaSC}), for the dressed
vertex in the superconducting state.}
\label{vertexSC}
\end{figure}

\begin{figure}
\epsfxsize=468bp
\epsfbox{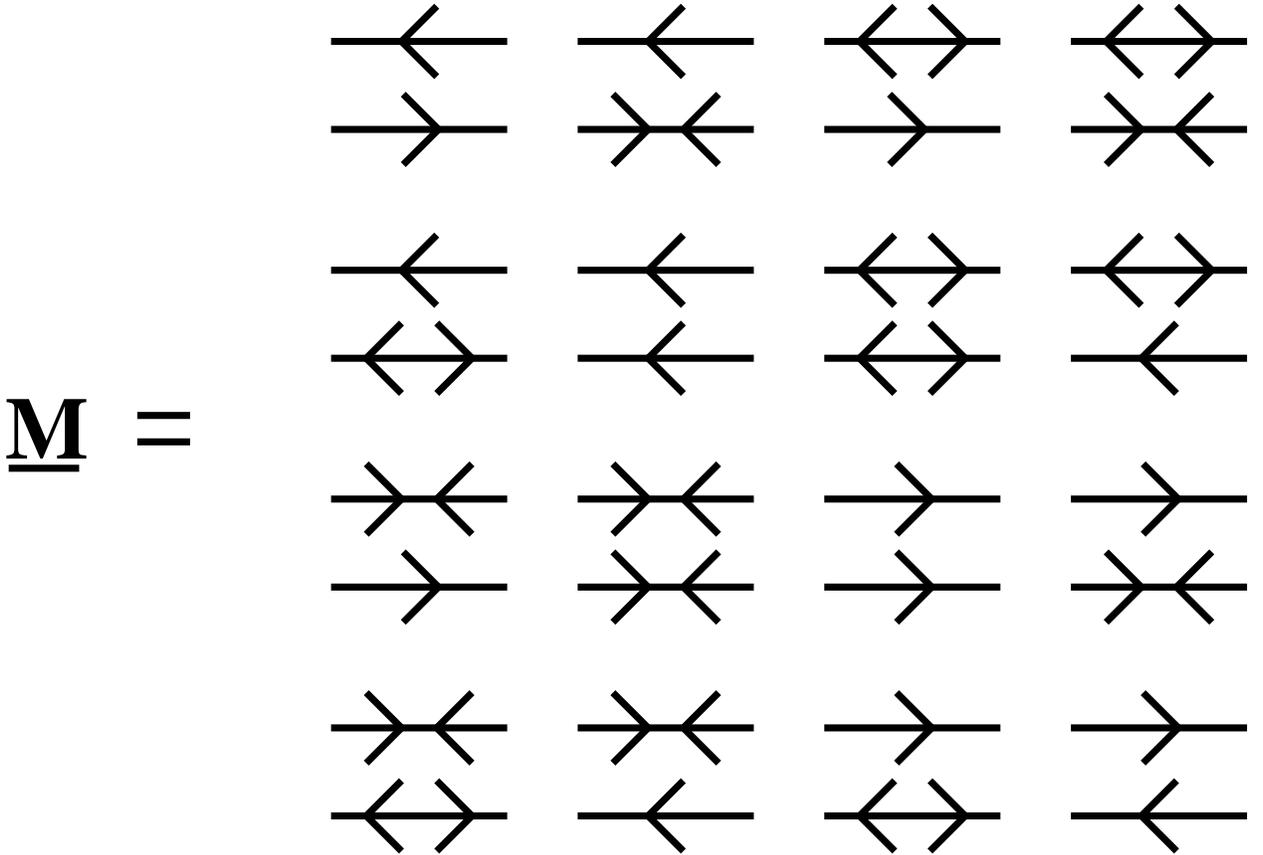}
\vspace{36bp}
\caption{Schematic representation of the diagrams corresponding to the
components of the matrix $\underline M({\bf q},i\omega_m,{i\omega_n})$.}
\label{Mmatrix}
\end{figure}

A complication which arises in the superconducting state is that the
dressed vertex cannot be represented as a simple scalar function.
Since the electron and hole propagators must be represented by both
normal and anomalous propagators to account for electrons scattering
into and out of the superconducting condensate, the dressed vertex is
represented by a four component vector which accounts for all possible
combinations of electron and hole lines going into and out of the
vertex, as shown in Fig.~\ref{vertcomponents}.
The equations for the
vertex corrected spin susceptibility

\begin{eqnarray}
\underline \chi({\bf q},i\omega_m) = T \sum_n ~ \underline M({\bf
q},i\omega_m,{i\omega_n}) ~
    \overline \Lambda({\bf q},i\omega_m,{i\omega_n})
\label{XvertSC}
\end{eqnarray}
and for the dressed vertex

\begin{eqnarray}
\overline \Lambda({\bf q},i\omega_m,{i\omega_n}) = \overline 1 +
\Gamma({i\omega_m},{i\omega_n}) ~
    \underline M({\bf q},i\omega_m,{i\omega_n}) ~ \overline
\Lambda({\bf q},i\omega_m,{i\omega_n})
\label{lambdaSC}
\end{eqnarray}
then become matrix equations where the vertices are 4-component
vectors (overscored symbols) and the particle-hole bubbles are
represented by 4x4 matrices (underscored symbols).  Equation
(\ref{lambdaSC}) represents four separate equations, one for each
component of $\overline \Lambda({\bf q},i\omega_m,{i\omega_n})$. The
diagrammatic representation of the first of these is shown in
Fig.~\ref{vertexSC}.  The diagrammatic representations for the other
three components are similar.  The diagrams corresponding to the
components of the matrix $\underline M({\bf q},i\omega_m,{i\omega_n})$
are shown schematically in Fig.~\ref{Mmatrix}.  Explicitly, the
components of $\underline M({\bf q},i\omega_m,{i\omega_n})$, which is
symmetric, are

\widetext
{\small
\begin{mathletters}
\begin{eqnarray}
\underline M_{11}({\bf q},{i\omega_m},{i\omega_n})
 &&= -\underline M_{22}({\bf q},{i\omega_m}+2{i\omega_n},-{i\omega_n})
   = -\underline M_{33}({\bf q},-{i\omega_m}-2{i\omega_n},{i\omega_n})
\nonumber\\
 &&= \underline M_{44}({\bf q},-{i\omega_m},-{i\omega_n})
   = - \int {d^2p \over (2 \pi)^2} ~
             G({\bf p}+{\bf q},{i\omega_m} + {i\omega_n}) ~
             G({\bf p},{i\omega_n}), \\
\nonumber\\
\underline M_{12}({\bf q},{i\omega_m},{i\omega_n})
 &&= \underline M_{13}({\bf q},-{i\omega_m},{i\omega_m}+{i\omega_n})
   = -\underline M_{24}({\bf q},{i\omega_m}+2{i\omega_n},
                        -{i\omega_m}-{i\omega_n})
\nonumber\\
 &&= -\underline M_{34}({\bf q},-{i\omega_m}-2{i\omega_n},{i\omega_n})
   = \int {d^2p \over (2 \pi)^2} ~
           G({\bf p}+{\bf q},{i\omega_m} + {i\omega_n}) ~
           F({\bf p},{i\omega_n}), \\
\nonumber\\
\underline M_{14}({\bf q},{i\omega_m},{i\omega_n})
 &&= \underline M_{23}({\bf q},i\omega_m,{i\omega_n})
   = - \int {d^2p \over (2 \pi)^2} ~
             F({\bf p}+{\bf q},{i\omega_m} + {i\omega_n}) ~
             F({\bf p},{i\omega_n}).
\end{eqnarray}
\end{mathletters}
}
\narrowtext
Just as for the normal state, Eqs.~(\ref{XvertSC}) and
(\ref{lambdaSC}) may be rearranged, yielding
\begin{eqnarray}
\underline \chi({\bf q},i\omega_m) = T \sum_n ~ \underline M({\bf
q},i\omega_m,{i\omega_n}) ~
   [\underline 1 - \Gamma({i\omega_m},{i\omega_n}) ~ \underline M({\bf
q},i\omega_m,{i\omega_n})]^{-1}.
\end{eqnarray}

As mentioned previously, the vertex representing magnetic neutron
scattering connects only electrons of opposite spin. Thus to recover
the quantity corresponding to the spin susceptibility measured by
magnetic neutron scattering we must now add the components of
$\underline \chi({\bf q},i\omega_m)$ which correspond to diagrams
which begin and end with opposite spin electron lines:
\begin{eqnarray}
\chi({\bf q},i\omega_m) = \underline \chi_{11}({\bf q},i\omega_m) +
\underline \chi_{41}({\bf q},i\omega_m).
\end{eqnarray}
To avoid double counting, $\underline \chi_{14}({\bf q},i\omega_m)$
and $\underline \chi_{44}({\bf q},i\omega_m)$ are not included.

\begin{figure}
\epsfxsize=396bp
\epsfbox{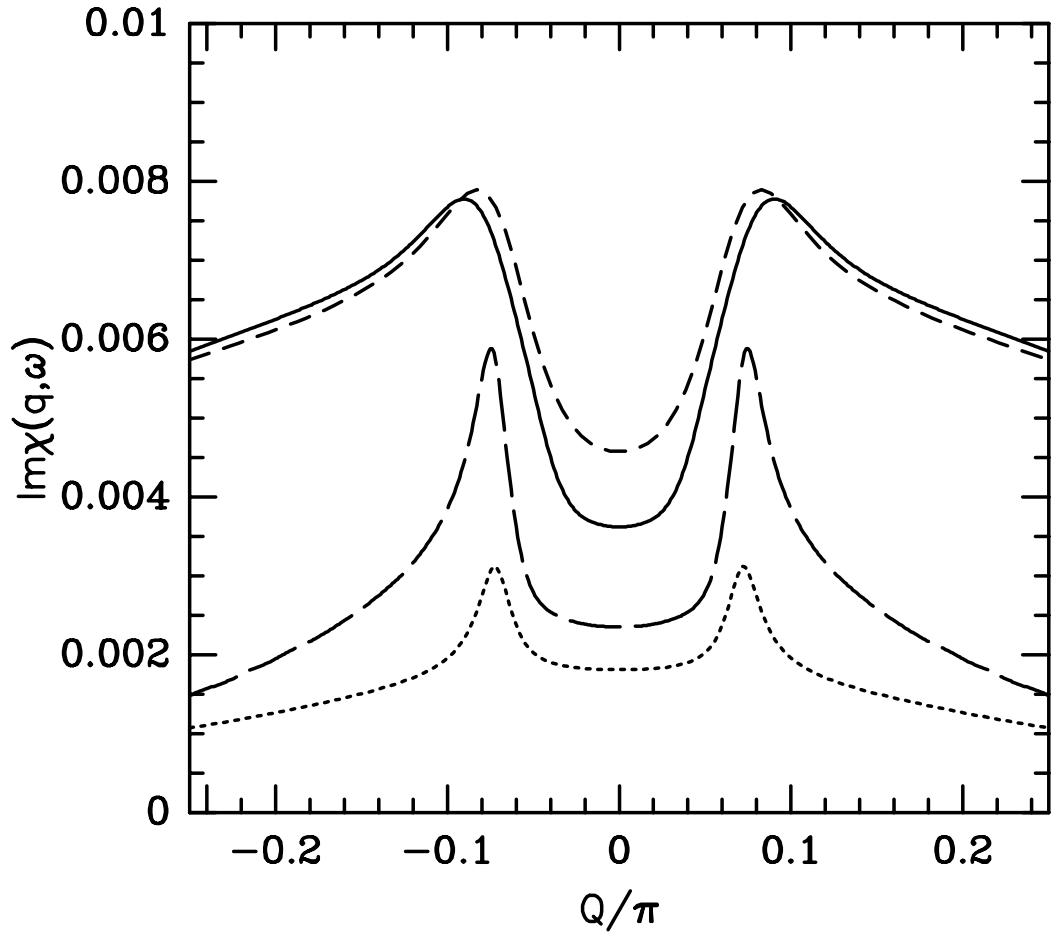}
\caption{$\chi''({\bf q},\omega)$ with and without the inclusion of
impurity vertex corrections as a function of momentum. The curves
represent the following: (solid) normal state, vertex corrected;
(short dashes) normal state, no vertex corrections; (long dashes)
superconducting state, vertex corrected; (dotted) superconducting
state, no vertex corrections.}
\label{vertVSnovert}
\end{figure}

Figure \ref{vertVSnovert} shows $\chi''({\bf q},\omega)$ calculated
with and without impurity vertex corrections.
Here $\chi({\bf
q},i\omega_m)$ has been calculated for imaginary frequencies. The real
frequency results were then obtained by analytic continuation using
Pad\'e approximants.\cite{paderefs} In the normal state the vertex
corrections do not have a dramatic effect on the zone edge peak
structures. There is a noticeable decrease in $\chi''({\bf q},\omega)$
along the zone diagonal. In the superconducting state the impurity
vertex corrections lead to a further enhancement of the peaks at the
zone edge positions. These peaks are now comparable in size to those
in the normal state.

\section{Coulomb interaction effects}
\label{coulombsection}

\begin{figure}
\epsfxsize=396bp
\epsfbox{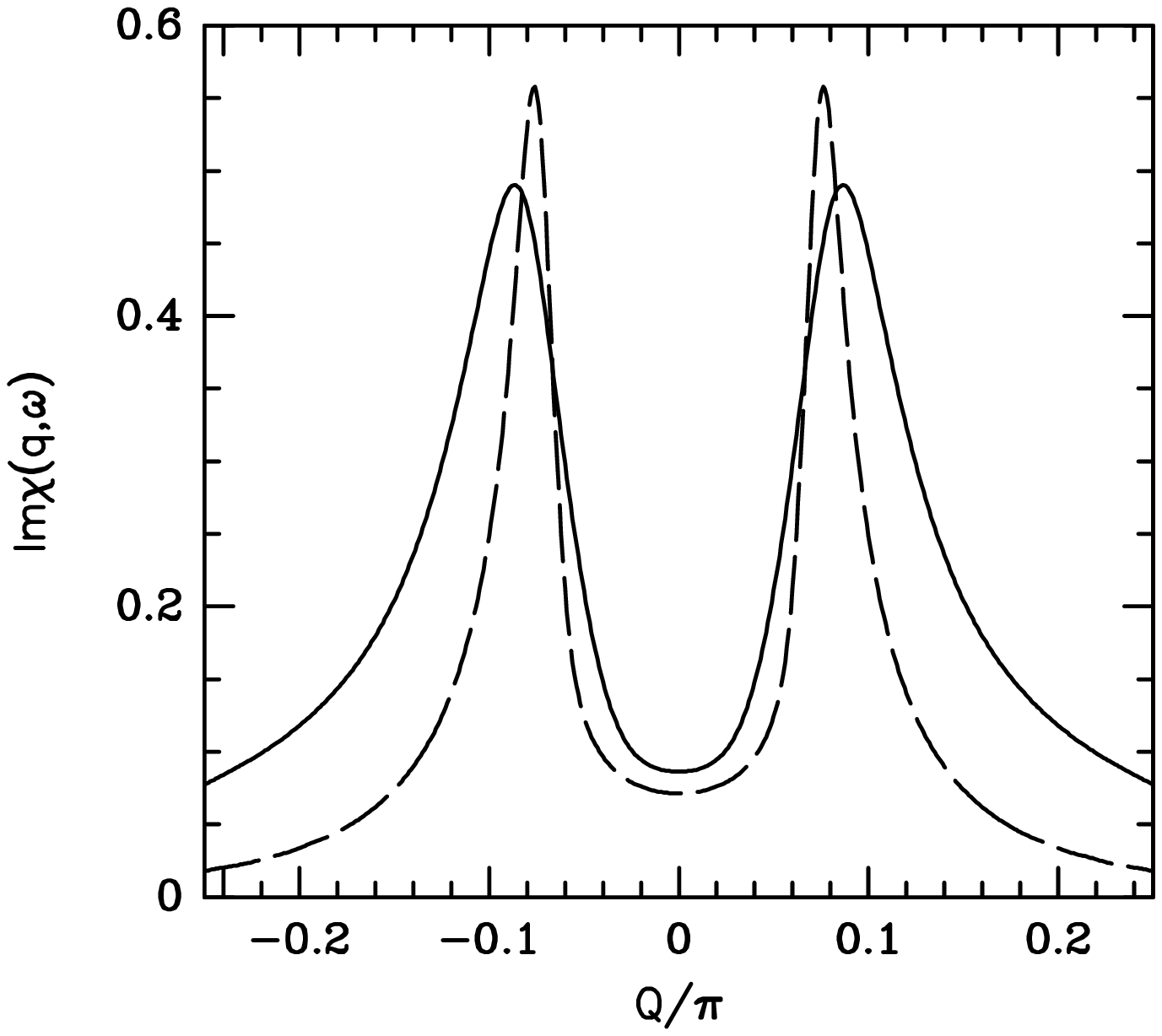}
\caption{$\chi''({\bf q},\omega)$ with Coulomb enhancement effects
included in the normal (solid line) and superconducting states (dashed
line) as a function of momentum.}
\label{chirpa}
\end{figure}

In addition to impurity scattering, it is important to take into
account the effects of the spin correlations produced by the Coulomb
interaction. These have been shown to play an essential role in
calculations of both the normal and the superconducting NMR
responses.\cite{kitaoka} Furthermore, as noted in the introduction,
Coulomb spin susceptibility enhancement effects have been shown to
maintain the neutron scattering peaks at their normal state positions
along the Brillouin zone edge.\cite{NejatXrpa,Maki} Just as in the NMR
and neutron scattering studies, we treat the Coulomb interaction
within an RPA approximation in which
\begin{eqnarray}
\chi({\bf q},\omega) = {\chi_0({\bf q},\omega) \over {1 - \overline U
\chi_0({\bf q},\omega)}}.
\end{eqnarray}
Here, however, $\chi_0({\bf q},\omega)$ is the spin susceptibility in
the presence of impurities discussed in Sec.\ref{impuritysection}. The
interaction $\overline U$ represents a renormalized Coulomb
interaction.\cite{BulutScalapinoWhite} Figure \ref{chirpa} shows
$\chi''({\bf q},\omega)$ calculated for $U=2t$.
The Coulomb
interaction enhances the normal state peak height by close to a factor
of 60. This results in peak heights for the model which are of the
same order of magnitude as the spin susceptibility peak heights
reported by Mason and
co-workers.\cite{Mason,masonpreprint,quinlanthesis}

As was mentioned in Sec.~\ref{impuritysection}, the Coulomb
interaction also gives rise to dynamic lifetime effects through
spin-fluctuation scattering. Previous calculations\cite{pines} have
shown that scattering of electrons by spin fluctuations leads to the
right order of magnitude for the quasiparticle relaxation rate in the
normal state of the cuprates, i.e., $\tau^{-1}(T_c) \sim T_c$.  In the
superconducting state, however, the gap in the electron excitation
spectrum leads to a suppression of the low energy part of the spin
fluctuation spectral weight and therefore to a suppression of the spin
fluctuation scattering contribution to the quasiparticle relaxation
rate.\cite{nuss} Previous calculations\cite{quinlantau} for a model
system similar to the one used here have shown that for $T=0.1T_c$ and
quasiparticle frequencies of order the temperature the quasiparticle
relaxation rate due to spin fluctuation scattering is suppressed by
several orders of magnitude compared to the value at $T_c$. For higher
frequencies, such as those used in the experiments of Mason and
co-workers, this relaxation rate is still small compared to the
impurity scattering rate for the concentration of impurities we have
considered here.\cite{quinlanthesis} Thus the impurity scattering
represents the dominant lifetime effect in the superconducting state
results discussed here. In the normal state, on the other hand, the
lifetime is determined by inelastic spin-fluctuation scattering.
However, in the normal state, lifetime effects lead only to small
changes in peak structure of $\chi''({\bf q},\omega)$, as shown in
Fig.~\ref{pureVSimp}.

\begin{figure}
\epsfxsize=396bp
\epsfbox{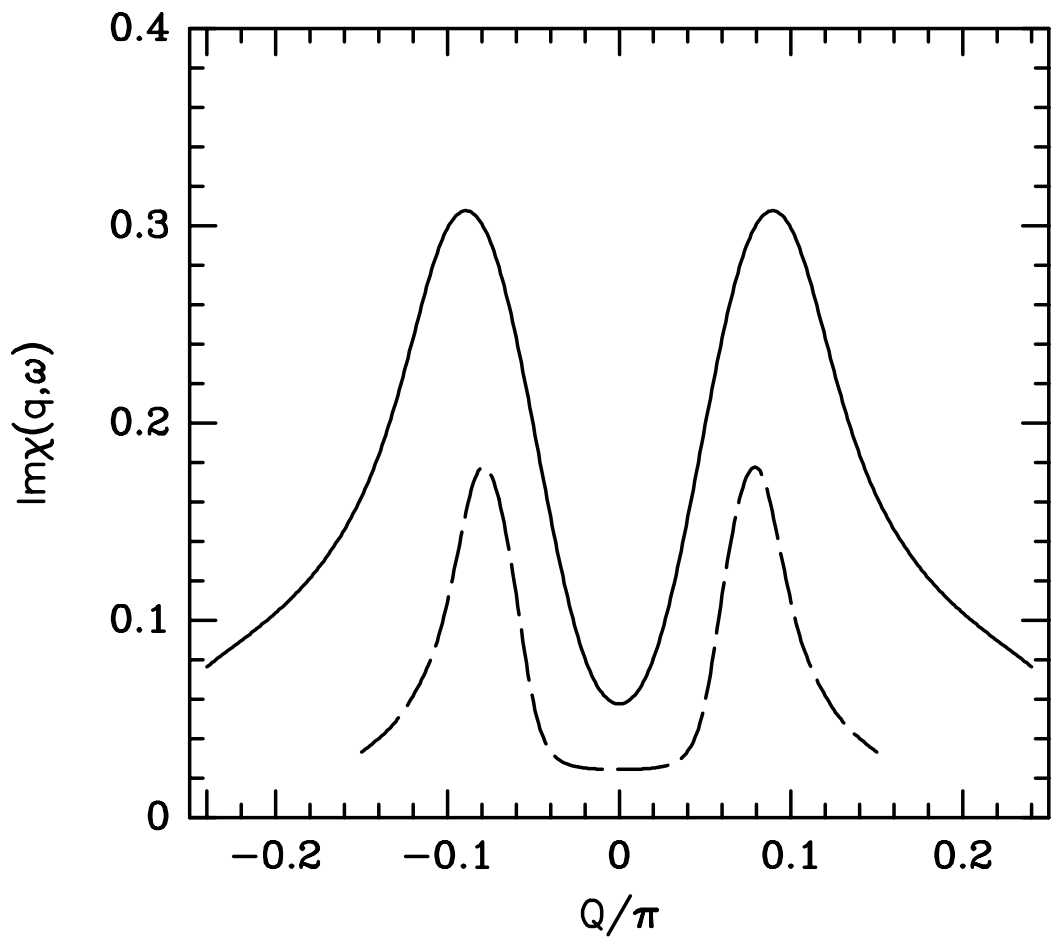}
\caption{Momentum averaged $\chi''({\bf q},\omega)$ in the normal
(solid line) and superconducting states (dashed line) as a function of
momentum.}
\label{momentavg}
\end{figure}

In order to directly compare the calculation including Coulomb
enhancement effects to the experimental results we must average
$\chi''({\bf q},\omega)$ over a region of momentum space comparable to
the experimental resolution.  Figure \ref{momentavg} shows
$\chi''({\bf q},\omega)$ averaged over a region in momentum space with
dimensions $0.04\pi$ by $0.12\pi$, the larger dimension being
perpendicular to the momentum scan direction.
These results show some
qualitative agreement with the experimental data. At the zone edge and
zone diagonal the magnitude of $\chi''({\bf q},\omega)$ is reduced by
a factor of about one half in the superconducting state compared to
the normal state. This reduction is, however, not quite isotropic. The
zone edge peak structures are noticeably narrower in the
superconducting state compared to the normal state.

\section{Conclusions}
\label{conclusions}

We have calculated the spin susceptibility at and below $T_c$ for a
model $d_{x^2-y^2}$--wave superconductor with resonant impurity
scattering and Coulomb correlations.  Impurity scattering effects,
including vertex corrections, act to restore the zone edge peaks in
the spin susceptibility which are otherwise suppressed by a $d$--wave
superconducting gap.  The predicted amount of suppression of the spin
susceptibility in the superconducting state compared to the normal
state is in qualitative agreement with the experimental results in
that the predicted neutron scattering intensity at the zone edge and
along the diagonal is suppressed by about one half in the
superconducting state compared to the normal state.  The peak
structure predicted by this model differs from the experimental
results in that the peaks narrow appreciably at low temperatures
compared to the normal state.

\acknowledgments

The authors would like to acknowledge insightful discussions with
N.~Bulut and P.~Monthoux. This work was partially supported by the
National Science Foundation under grants DMR92-25027. The numerical
calculations reported in this paper were performed at the San Diego
Supercomputer Center and the National Energy Research Supercomputer
Center.

\end{document}